\documentclass[12pt]{article}
\usepackage{geometry}               
\usepackage{graphicx}
\usepackage{amssymb}
\usepackage{amsmath}
\usepackage{epstopdf}
\usepackage{float}
\usepackage[small]{caption}
\usepackage{epstopdf}  

\begin{document}

\begin{titlepage}
\vfill

\begin{flushleft}
{\Large\bf 3-dimensional structure of a sheet crumpled into a ball$^*$}

\end{flushleft} 
\begin{flushleft}
\vskip 1.0cm
{\large {\sl }}
\vskip 10.mm
{\bf Anne Dominique Cambou and Narayanan Menon} \\
\vskip 0.5cm
{
Department of Physics\\
University of Massachusetts\\
Amherst, MA 01003\\
USA\\

E-mail: menon@physics.umass.edu

     	     }
\end{flushleft}
\vskip 0.5in
\begin{flushleft}
{\bf{Abstract.}}  When a thin sheet is crushed into a small three-dimensional volume, it invariably forms a structure with a low volume fraction but high resistance to further compression.  Being a far-from-equilibrium process, forced crumpling is not necessarily amenable to a statistical description in which the parameters of the initially flat sheet and the final confinement fully specify the resulting crumpled state. Instead, the internal geometry and mechanical properties of the crumpled ball may reflect the history of its preparation. Our X-ray microtomography experiments reveal that the internal 3-dimensional geometry of a crumpled ball is in many respects isotropic and homogeneous. In these respects, crumpling recapitulates other classic nonequilibrium problems such as turbulence, where a system driven by long-wavelength, low-symmetry, forcing shows only rather subtle fingerprints of the forcing mechanism. However, we find local nematic ordering of the sheet into parallel stacks. The layering proceeds radially inwards from the outer surface. The extent of this layering increases with the volume fraction, or degree of compression.

\end{flushleft}

\end{titlepage}

While the global 3-dimensional arrangement of a crumpled object is very complex, a remarkable feature of the geometry is that a large fraction of the surface area is relatively flat. The curvature imposed on an elastic sheet by external confinement or forces dwells is concentrated largely in a network of ridges \cite{ alexlobkovsky,witten} which meet at vertices known as developable cones \cite{cerda,chaieb}. For a perfectly elastic sheet, the work done in crumpling is stored in the elastic energy of these focused deformations, which is partitioned in finite fractions of bending and stretching energies \cite{alexlobkovsky}. For most familiar examples of crumpled sheets such as plastic, paper, or metal foils, the strains at ridges exceed the yield strain and the ridges become irreversibly creased into folds. In the crumpled regime, these focused structures must interact simply due to geometric confinement. Understanding the mechanics of an interacting set of folds is a formidable challenge, and it is crucial to obtain experimental insights into their 3-dimensional arrangement.

Some important lessons are learned by studying a lower-dimensional version of the problem, that is, the packing of a 1-dimensional
elastic curve confined in 2-dimensions. Experiments on the packing of a wire \cite{donato} in a circular cage, and of sheets
pulled through a circular hole \cite{deboeuf} showed a new feature: as a curve is confined, it starts organizing into
parallel arcs. Based on a recent lattice simulation of this problem \cite{Radin}, it has been argued that true long-range
nematic order can be obtained in the thermodynamic limit.  In this lower dimensional problem, it is possible for the curve to pack without
any stretching, and therefore, no focused structures are produced even in the limit of zero thickness. However, 2-dimensional
sheets confined in a spherical space are subject to greater geometric frustration - they cannot accommodate the constraints
presented by confinement in all three directions solely by bending. They are forced to stretch and develop gaussian curvature by bending
in two directions \cite{venkataramani}. It is thus important to come to terms with the fully 3-dimensional problem as there are qualitative differences between these situations.

Two recent sets of numerical simulations of purely elastic \cite{vliegenthart} and elasto-plastic \cite{tallinen} self-avoiding
sheets in 3-dimensions have studied the distribution of elastic energy in ridges and the scaling of compressive forces with confinement.
The configurations shown in Ref. \cite{tallinen} clearly indicate that layering also occurs in 3D. While there is no direct
spatial quantification of this phenomenon, they find that the energy becomes dominated by self-contacts as the confinement increases.

It is  a challenge to experimentally probe the interior geometry of a 3D crumpled object. Ref. \cite{blair} studied by laser profilometry the statistics of folds and vertices in an unfolded crumpled sheet. However,  unfolding the sheet leads to a loss of spatial information about the interactions of the folds and of the sheet's final crumpled configuration. A recent paper \cite{Aharony} studied by laser sheet illumination the early stages of energy focusing in an elastic sheet. X-ray tomography is a tool that is well-suited for providing structural data in the highly crumpled regime.  In recent X-ray tomography experiments \cite{linPRL09, linPRE09} a 2D analysis of the geometry was performed, in terms of tangent vectors to the curves obtained in 2-dimensional slices of the crumpled object.  This analysis showed parallel orientation of tangent vectors; thus both simulations and experiment indicate that layering of facets is a geometric element to be considered alongside focus!
 ed structures such as ridges and vertices. However, a truly 3-dimensional experimental characterization of the highly-crumpled state remains to be done.

In this article we present measurements by X-ray microtomography of the location of a crumpled sheet in a 3D volume, along
with fully 3-dimensional analyses of the curvature, orientation and layering of the sheet. We concentrate on an analysis
the radial dependence of these metrics of the geometry, in order to identify any possible anisotropy and inhomogeneity imposed
on the sheet by the method of crumpling.

\begin{figure} [H]
\includegraphics[width=\textwidth]{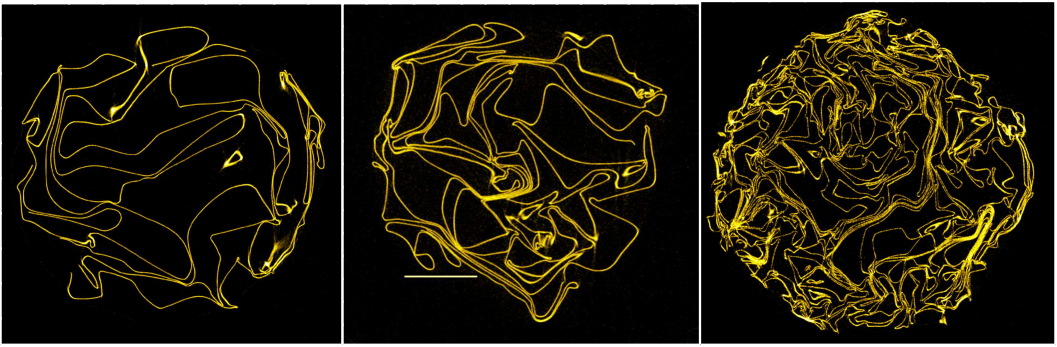} 
\caption{\label{fig:slices} Reconstructed slices through an equatorial plane of three crumpled spheres
with average volume fractions of $\phi=$ 6\%  ($R_{i} \approx$ 3.5cm, $R_{o}\approx$ 0.72cm), 8.5\% $R_{i}\approx$ 5cm, $R_{o}\approx$ 0.82cm) and 22\% ($R_{i}\approx$ 7cm, $R_{o}\approx$ 0.75cm), respectively. To remove background noise, all reconstructed images are thresholded before analysis.}
\end{figure}

In our experiments, circular aluminum sheets with thickness $t = 25.4\mu{m}$
and initial radius, $R_{i}$ are hand-crumpled into approximately spherical balls with final radius $R_{o}$. We have studied
balls with average volume fractions of  $\phi$ $\approx {6\%}$, $8.5\%$ and $22\%$. X-ray computerized tomography (CT) is used to image the interior of the crumpled sheet. A CT scanner (Skyscan 1172) images the radiation from a divergent x-ray illumination from a point source.
 Transmission images are taken after 0.3 degrees rotations of the sample over a total of 180 degrees. These images  are
 reconstructed into a three-dimensional stack of gray scale images with a typical size of $\sim{2000 \times 2000 \times 2000}$
 pixels. The voxels are uniformly distributed in the volume, and are cubes with a linear dimension of $8.85 \mu m \approx t/3$, thus ensuring that the thickness of the sheet is fully resolved.  As can be seen in the 2D slices through the crumpled balls shown in Fig. \ref{fig:slices}, we are now in a position to extract various quantitative measures of the disposition of the sheet in the crumpled volume.

\section{Results and Analysis}

\subsection{Mass Distribution}

\begin{figure} [H]
\includegraphics*[width=0.9\textwidth]{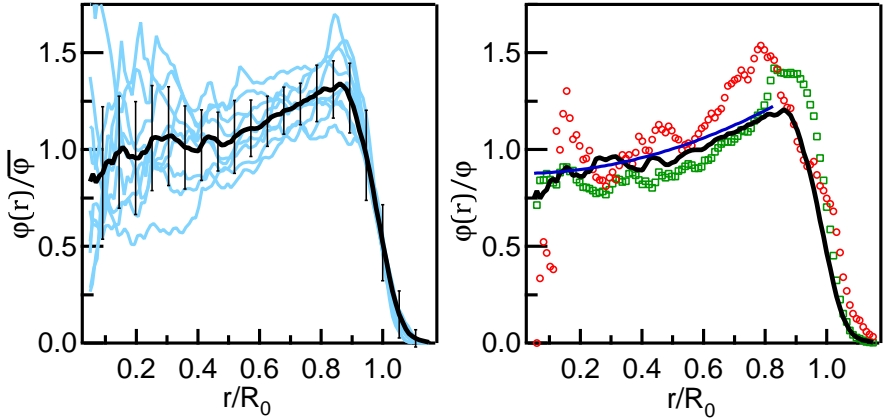}
\caption{\label{fig:avgmassdist} Mass Distribution. \emph{A (left), Radial dependence of volume fraction, $\phi(r)$ versus $r/R_o$, the radial distance $r$ from the center of mass normalized by $R_o$, the radius of the ball. The center of the sphere is defined by its center of mass. The data shown are for spheres with a radially averaged volume fraction of $\phi$ = 8.5\%.  The volume fractions are normalized by $\bar{\phi}$, where the over-bar represents a further average over samples. The blue curves are for nine separate samples, to give an indication of the sample-to-sample variability. The black curve is the average of these samples. Error bars indicate the standard deviation. The average mass distribution increases $\approx 30\%$ from the centre of the sphere to the exterior. B (right), Normalized volume fraction $\phi(r)/\phi$, versus the radial distance from the center of mass for spheres with $\phi$ = 6\%, 8.5\%, and 22\%. The red open circles and green open squares corres!
 pond to spheres with $\phi$ = 6\% and 22\% respectively. The black curve is the average of 9 spheres with $\phi$ = 8.5\%. The solid line is a best-fit to the equation $\phi_{fit} = \frac{\phi}{1+3C/5}(1+Cr^2/R_0^2)$,  with an adjustable parameter $C$.}}
\end{figure}

The simplest representation of the geometry of the sheet is the radial dependence of the volume fraction $\phi(r)$. We ensure by calibrating with a sample of known geometry that the mass of film determined from the reconstructed images is not affected by x-ray absorption through the volume of the sphere (see SI).  As the individual blue curves in Fig. \ref{fig:avgmassdist}A show for nine different crumpled sheets with the same average volume fraction $\phi=8.5\%$, there is considerable variation of $\phi(r)$ from sample to sample. The black curve, which is the average over these samples,  more clearly shows a trend of volume fraction increasing from the interior to the exterior of the sphere.  Thus the most elementary analysis reveals a signature of the low-symmetry route to the final crumpled state, where the confining forces are radial and inwards. The large variability within any of the individual samples reflects the very heterogenous distribution of void space with the ball. This heterogeneous distribution has often been quantified in terms of a fractal distribution of mass \cite{linPRE09, gomes, balankin2007a, balankin2007b}. However, given that there is an overall radial density gradient, a fractal dimension computed from an average over the volume of the ball is not a useful measure of heterogeneity.
The data shown in Fig. \ref{fig:avgmassdist}B compare the radial gradient of volume fraction for three different degrees of confinement. The functional dependence on radial distance does not change significantly with volume fraction over this range of $\phi$. This must of course change at extremely low and high \cite{linPRE09} confinement where a more homogeneous distribution might be expected.

\subsection{Orientation}

\begin{figure} [H]
\includegraphics*[width=0.9\textwidth]{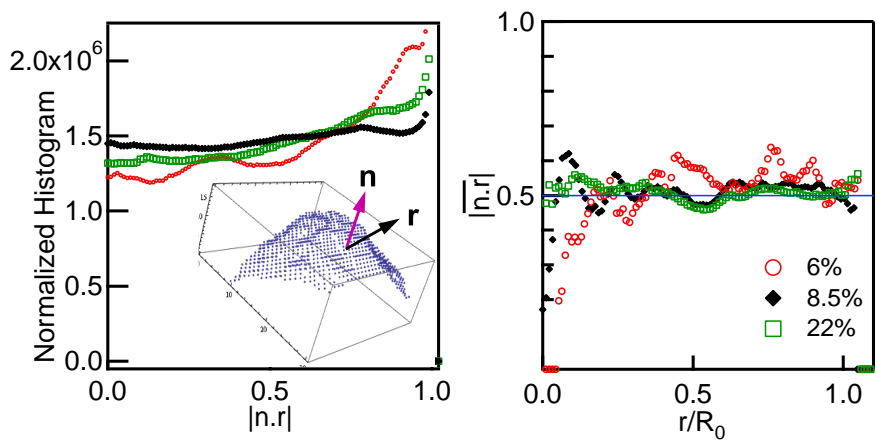}
\caption{\label{fig:orienthist} Orientation.\emph{ We quantify local orientation of the sheet by the dot product $\hat{n}\cdot\hat{r}$ of the surface normal and the radial unit vector from the centre of the ball, as shown in the 3D patch of sheet on the left. We display orientation data from spheres with $\phi$ = 6\%, 8.5\%, and 22\% (red open circles, black diamonds, and green open squares respectively). The 8.5\% data are an average of 2 spheres. A (left),Histogram of the sheet orientation, $\left |\hat{n}\cdot\hat{r}\right |$. All three volume fractions have a nearly isotropic orientation with only a slight tendency for alignment of the normals with the radial vector.  B (right), Average orientation
versus distance from the center of mass for $\phi$ = 6\%, 8.5\%, and 22\%. The solid line displays the average of an
isotropic distribution of orientations.}}
\end{figure}

We next move from the location of the sheet, to the orientation of sheet within the volume. From the gray-scale 3D image, we find the surface normal $\hat{n}$ at all points along the surface. The spatial orientation of the surface normal is quantified by the direction cosine $\hat{n} \cdot \hat{r}$, where $\hat{r}$ is the unit vector from the centre of the sphere to the surface point. The distribution of this direction cosine, shown in Fig. \ref{fig:orienthist}A is nearly uniform, thus indicating a near-isotropic distribution of sheet normals. This is a surprising result, given that the crumpling process might be expected to break symmetry between radial and azimuthal directions, and perhaps favour orientation of the sheets parallel to the confining surface, in onion-like fashion. This expectation of alignment by the outer surface is not borne out even as one approaches the surface. In Fig. \ref{fig:orienthist}B, we plot the average $\hat{n}\cdot\hat{r}$ as a function of normalized radial position $r/R_o$.  The orientation, $\hat{n}\cdot\hat{r}$, does not deviate from the value expected for random orientation, except perhaps within a small region close to $R=R_o$.  Except for this boundary layer, the orientation of surface normals is both isotropic and homogeneous within the volume, despite the sheet's route to the crumpled state.

\subsection{Curvature}

\begin{figure} [H]
\centering
\includegraphics*[width=0.9\textwidth]{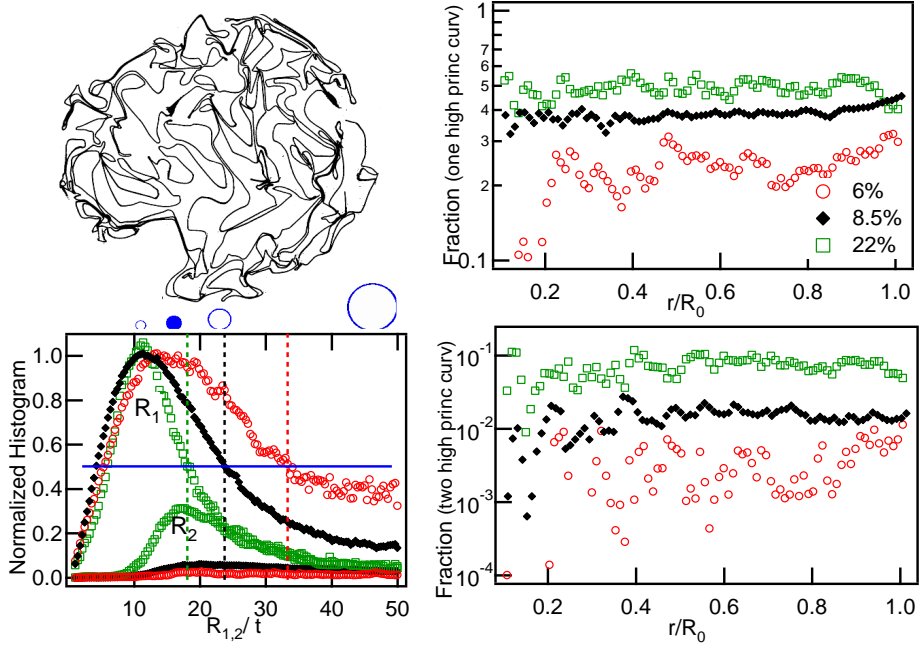}
  \caption{\label{fig:curvhist} Curvature. \emph{Displayed are measurements of the high curvature regions of spheres with $\phi$=6, 8.5\% and 22\%  (red open
  circles, black diamonds, and green open squares respectively). The 8.5\% data are an average of 2 spheres. A (top left). A reconstructed  image from a crumpled ball with $\phi$=8.5\% is shown to give a sense of scale for the radii of curvature.  B (bottom left).  Normalized histogram of the larger and smaller principal curvature-radii $R_1$ and $R_2$ in units of $t$. Radii of curvature between $3t$ and $50t$ are shown. There is a peak in the histogram at $R_1/t \approx 10$ for all three $\phi$. Locations with radius of curvature below a cutoff radius, $r_c$, are designated as high-curvature regions;  $r_c$ is chosen to be the radius at which the normalized histogram falls to half its peak value and its value is indicated for each volume fraction by the dashed vertical line. From left to right, the three open circles show the radius at the peak in $R_1$ ($10t$),  $r_c$ for 8.5\% ($24t$), and the upper limit of curvature measurements ($50t$). C (top right). The fraction of s!
 urface points with at least one high-curvature direction (folds and vertices) versus distance from center of mass. D (bottom right) The fraction of surface points with two high-curvature directions (vertices only) versus distance from center of mass. The high-curvature regions are homogeneously distributed.}}
\end{figure}

The next higher order in the description of the local geometry of the sheet is the curvature of the sheet. In principle, the
curvature can be determined from a complete knowledge of the vector field of surface normals $\hat{n}$ \cite{kamien}. However, numerically finding the gradient of this field introduces undesired inaccuracy. We chose to make an independent measurement of the curvature by identifying connected patches of the sheet, and fitting ellipsoids to these patches. From the fit parameters, we were able to obtain the two principal radii of curvature $R_1$ and $R_2 (>R_1) $ at every point on the patch. The unusual geometry of
the crumpled state presents a technical challenge here. While there are many regions of low curvature, the curvature can change sharply at the stress-condensed regions. Thus the surface has to be fit with rather small patches. It was not possible to infer large radii of curvature with great numerical accuracy by fitting to small patches in space. We therefore focus our analysis on highly-curved regions with radii of curvature $<50t$, which can be reliably fit.  In Fig. \ref{fig:curvhist}B, we show for three volume fractions, the high-curvature part of the histogram of the principal radii of curvature. There is a peak in the histogram at $R_1$ $\approx 10t$, independent of volume fraction. The radius associated with the peak in the histogram, $R_1=10t$, and with the largest fitted radii $R_{1,2}=50t$, are indicated by open circles in Fig. \ref{fig:curvhist}A. The value of curvature radius at the peak is much higher than that required to introduce plastic folds in the aluminum. The yield stress for Al is $145 MPa$ \cite{AlHandbook}, and a radius of curvature for yielding of $250t$ may be inferred from the corresponding yield strain. It thus appears that there is a limiting value of curvature beyond which the crumpling process creates new features, rather than compressing existing folds to sharper dihedral angles.  As we discuss below, though, the peak at small radius of curvature does not imply a high level of energy condensation. (Also see SI for further details).

From the curvatures we have determined, we identify surface points with one and with both radii of curvature below a cutoff radius,
and associate these with folds and vertices, respectively. The cutoff radius, $r_c$, for each $\phi$ is chosen to be the radius at which the histogram falls to half the value of the peak in the curvature distribution, as shown by the dotted lines in Fig. \ref{fig:curvhist}B. The number of high-curvature points identified by applying an arbitrary threshold obviously depends on the value of $r_c$. With the definition we adopt, we find that the fraction of surface points associated with folds is 37\% and with vertices is 4\% for the $\phi$=8.5\% sample (Fig. \ref{fig:curvhist}C,D). The large fraction of surface points with small radii of curvature indicates that in this regime of confinement, stress condensation is incomplete and that it is not correct to think of the sheet as largely flat. In Fig. \ref{fig:curvhist}C we display the fraction of points with $R_1< r_c$ (folds and vertices), as a function of radial position. Fig. \ref{fig:curvhist}D shows the same quantity for points with
$R_1<R_2<r_c$ (vertices). It is evident from these data that the high-curvature parts of the sheets are homogeneously distributed through the sphere, and show no evidence of being preferentially generated at the confining walls. A plausible explanation for the
radial increase in volume fraction might have been that the exterior part of the sphere allows more gentle curvature, and therefore that it is energetically preferred to have more of the sheet reside in the exterior. In light of the observation of a homogeneous distribution of high-curvature features, and of the isotropic orientation of the sheet normals, this explanation is not tenable. This homogeneous curvature distribution contrasts with crumpling  in 2-dimensions \cite{deboeuf}, where curvature is different in the bulk and at the boundary, where the object assumes the curvature of the confining wall. Possibly this is due to the fact that the spherical boundary condition imposes a gaussian curvature on a flat sheet, which unlike in the 2D analogue, is not possible to accommodate without stretching deformations.

\subsection{Nematic Ordering}

\begin{figure} [H]
\centering
\includegraphics*[width=0.9\textwidth]{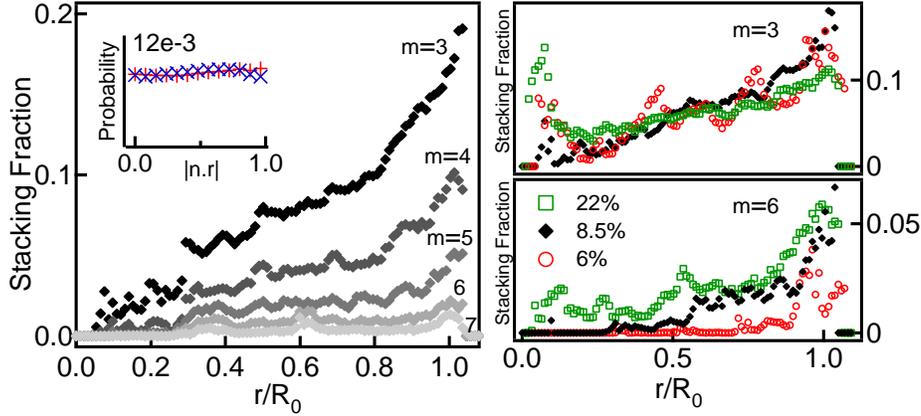}
\caption{\label{fig:stack} Stacking. \emph{Stacks are identified by starting at all surface points on the sheet,
searching in the direction of the normal $\hat{n}$, and identifying the number, $m$, of other surfaces with anti-parallel normals within a radius of $16t$. For example, $m$=1 represents an isolated part of the sheet, and $m$=4 represents 4 closely-stacked surfaces. The solid circle in Fig. \ref{fig:curvhist} gives a visual sense of the search radius.  A (left). Fraction of surface points participating in stacking versus distance from the centre of mass for different values of $m$ in one sphere with $\phi$ = 8.5\%. The symbols represent stacks containing $m$=3,4,5, 6 and 7 facets (darkest to lightest colored diamonds respectively). The stacking fraction increases with $r/R_o$ for all values of $m$. Inset, histogram of surface orientation, $|\hat{n} \cdot \hat{r}|$, for points on stacked surfaces (i.e. $m\geq$2) only (blue crosses). We also show for comparison the histogram for all surface points (red plusses). This shows that the stacked facets are also oriented isotropically!
 . In the panels on the right we show stacking fraction versus $r/R_o$ for $\phi = 6\%, 8.5\%$,and $22\%$ (red open circles, black diamonds, and green open squares, respectively) for  (B) $m$=3 and   (C) $m$=6. As $\phi$ increases,  stacks become thicker.}}
\end{figure}

Thus far, we have concentrated entirely on \emph{local} descriptors of the geometry. We now turn to the stacking of facets that is
 evident in Fig. \ref{fig:slices}.
 As a quantitative measure of this local nematic ordering, we study the correlation of the 3D surface-normals along the sheet. From each surface point we search along the normal direction for other surfaces in parallel alignment. We label surfaces stacked together in this way within a chosen search radius, and count the number of sheets layered together in each stack, $m$. In Fig. \ref{fig:stack}A, we show as a function of radial position, $r/R_o$, the fraction of surface points which are participating in stacks composed of $m=$3 to 7. These data for $\phi$=8.5\% reveal a tendency to stack that increases approximately linearly in the radial direction. In Fig. \ref{fig:stack}B we show the probability of occurrence of $3$-stacks and $6$-stacks for different volume fractions. At all three volume fractions
 we study, the probability of $3$-stacks is similar, and increases approximately linearly in the radial direction. However, $6$-stacks are just beginning to form in the outer layers of the sphere at lower volume fractions, and develop more strongly at higher volume fractions. The implication is that stacking is initiated from the outside and progresses inwards as crumpling proceeds. Thicker stacks could then be produced by accretion or folding of thinner stacks, and are not due to a strong increase in the overall fraction of layered sheet.

As mentioned earlier, the crumpled state spontaneously develops structural rigidity at very low volume fractions without externally
imposed design. This rigidity has been attributed to the formation of ridges with high buckling strengths\cite{vliegenthart, tallinen}. While there is no direct mechanical verification of either mechanism, the layering shown in Fig. \ref{fig:stack} may also be a contributing factor to
mechanical rigidity. That is, the structure may be stabilized against external compression by forming multi-layered walls rather than pillars. The force threshold to buckle a ridge into smaller ridges\cite{DiDonna}, or for Euler buckling of a planar region, scales with the bending modulus of the sheet, which grows as $t^3$, the cube of its thickness. If the sheets in an $m$-stack do not slide relative to each other either due to frictional contact, or by the ridges that might delimit the edges of the layered region, then the rigidity of such a stack scales as $(mt)^3$. Stacking can thus greatly enhance mechanical rigidity.
The relationship between stacking and stress condensation remains to be clarified. It has been suggested in studies of phyllotaxy in cabbages that layered arrangements are nucleated between stiff leaf stems \cite{CouturierThesis}.
A similar mechanism may be operative here, with the folds playing the role of the stem, and forcing sheets in near-contact to make stacks. Even though the stacks are formed under radial compression they are also isotropically arranged, and potentially strengthen the structure against forces in any direction. This is shown in the inset to Fig. \ref{fig:stack}A where we separately display the distribution of orientation with and without the stacked layers. (A different result was obtained when a 2D determination of orientation was performed \cite{linPRL09}). Any arbitrary force applied to the crumpled ball will result in both compression and shear internally. The response of the stack to such a generalized stress state may involve both the self-avoidance of the sheet as well as friction between the sheets.

The detailed geometrical characterization of the crumpled state raises questions
that we are currently attempting to address by simultaneous studies of structure and dynamics. However,
the principal inference to be drawn from the structural information presented here is that several
aspects of the geometry are to a good approximation homogeneous and isotropic, and therefore perhaps
amenable to statistical treatment.  Despite the low-symmetry path to the crumpled state, it is rather
remarkable that from the vantage point of a location in the interior of the ball, no local measurement
of geometry points the way to the exterior of the ball.

\subsection{Methods and materials}
To find the radial distribution of mass, the reconstructed image stack is first thresholded, and then despeckled to remove detector
noise. $\phi(r)$ is determined by counting occupied voxels in spherical shells of thickness $\Delta{r}=3.3t$. We ensure
by measurements on a calibration sample that absorption effects do not affect the measured $\phi(r)$ (see SI). We choose to
define the radial distance $r$ from the centre of mass of the ball. The exterior hull of the ball is not perfectly spherical, nor
is the centre of mass necessarily at the geometric centre of this hull. However, at these average volume fractions, the differences
between these measures is not significant for the results we discuss.  The final radius $R_o$ is determined from the steep fall-off
in $\phi(r)$ at the exterior of the ball, and is chosen to be where  $\phi(r) \approx$ 0.045$\phi$. $R_o$ is
consistent with measurement of the caliper radius of the crumpled ball, averaged over orientation.

The orientation of the sheet is determined by applying a 3D canny edge-detector to find the surface normal $\hat{n}$ at all points. The edges are smoothed by a Gaussian filter. We use a simple threshold to fill in low-intensity surface points.

To measure curvature, the sheet is first thinned by identifying the number of neighbors in a $3\times3$ volume surrounding that point. This procedure reduces both surfaces of the sheet to a one-voxel-thick edge. All detected surface points are only counted once. The entire sheet surface is decomposed into connected patches that are fitted to algebraic functions (see SI) from which the local curvatures are determined. Our analysis is optimized toward accurate determinations of the high-curvature features of the sheet.

We identify the number of facets stacked between two parallel surfaces by searching in the  direction of the normal at each surface point, $\hat{n_1}$, at a radius of up to $\sim 16t$, (see the solid circle below Fig. \ref{fig:curvhist}A for an visual indication of the search radius). We identify all other surface points along this search direction with normals $\hat{n_2}$ that are antiparallel to $\hat{n_1}$ as defined by the criterion $\hat{n_1}\cdot\hat{n_2}<-0.97$.  We then return to the raw image to count all occupied voxels between these points with antiparallel normals. $m$-stacks are identified by dividing this count by the sheet thickness measured in pixels. When the sheets are very tightly stacked, the  interior surfaces in contact are not resolved in the images. However, our identification of the number of sheets, $m$, in the stack is not affected by this complication, since we count the entire mass between the two outermost sheets in the stack.

We gratefully acknowledge Whitey Hagadorn and Diane Kelley for access and guidance to the microtomography apparatus. We also thank H-q. Wang, G.R. Farrell, and H. Aharoni for helpful conversations. Financial support for this work was provided through NSF-DMR 0907245 and the MRSEC at UMass Amherst.

(*)Published online before print August 22, 2011, doi: 10.1073/pnas.1019192108

\end{document}